\UseRawInputEncoding
\documentclass[lettersize,journal]{IEEEtran}
\usepackage{amsmath,amsfonts}
\usepackage{algorithmic}
\usepackage{algorithm}
\usepackage{amsmath}
\usepackage{array}
\usepackage[caption=false,font=normalsize,labelfont=sf,textfont=sf]{subfig}
\usepackage{textcomp}
\usepackage{stfloats}
\usepackage{url}
\usepackage{verbatim}
\usepackage{graphicx}
\usepackage{cite}
\usepackage{epsfig}
\usepackage{bm}
\usepackage{makecell}
\usepackage{multirow}
\begin{document}

\title{Low-complexity Resource Allocation for Uplink RSMA in Future 6G Wireless Networks}

\author{\thanks{Gang Liu's work is supported by NSFC project under Grant No. 61971359, Sichuan Science and Technology Program under Grant No. 2023ZHCG0010, No. 2023YFH0012 and No. 2023YFG0312. Zheng Ma's work is supported by NSFC projects No.62271419 and U2268201. The corresponding author is Gang Liu.}
Jiewen Hu\thanks{Jiewen Hu is currently pursuing the Ph.D. degree at the Provincial Key Lab of Information Coding and Transmission, Southwest Jiaotong University, Chengdu 611756, China. (email: jevon@my.swjtu.edu.cn).},
Gang Liu\thanks{Gang Liu is currently an associate professor at the Provincial Key Lab of Information Coding and Transmission, Southwest Jiaotong University, Chengdu 611756, China. (email: gangliu@swjtu.edu.cn).}, \emph{Member, IEEE},
Zheng Ma\thanks{Zheng Ma is currently a professor at the Provincial Key Lab of Information Coding and Transmission, Southwest Jiaotong University, Chengdu 611756, China. (zma@home.swjtu.edu.cn).}, \emph{Member, IEEE},
Ming Xiao\thanks{Ming Xiao is with the Department of Information, Science and Engineering,
School of Electrical Engineering, KTH, 10044 Stockholm, Sweden (e-mail:
mingx@kth.se).}, \emph{Senior Member, IEEE},
and Pingzhi Fan\thanks{Pingzhi Fan is currently a distinguished professor the Provincial Key Lab of Information Coding and Transmission, Southwest Jiaotong University, Chengdu 611756, China. (pzfan@swjtu.edu.cn).}, \emph{Fellow, IEEE}
}



\maketitle

\begin{abstract}

Uplink rate-splitting multiple access (RSMA) requires optimization of decoding order and power allocation, while decoding order is a discrete variable, and it is very complex to find the optimal decoding order if the number of users is large enough. This letter proposes a low-complexity user pairing-based resource allocation algorithm with the objective of minimizing the maximum latency. Closed-form expressions for power and bandwidth allocation for a given latency are first derived. Then a bisection method is used to determine the minimum latency and optimal resource allocation. Finally, the proposed algorithm is compared with unpaired RSMA using an exhaustive method to obtain the optimal decoding order, unpaired RSMA using a suboptimal decoding order, paired non-orthogonal multiple access (NOMA) and unpaired NOMA. The results show that our proposed algorithm outperforms NOMA and achieves similar performance to unpaired RSMA. In addition, the complexity of the proposed algorithm is significantly reduced.
\end{abstract}

\section{Introduction}
6G takes the upper bound of wireless access capacity to a new level, and it expects ubiquitous connectivity, which is overwhelming for existing systems. An excellent multiple access scheme can effectively solve the above problems, and rate-splitting multiple access (RSMA) is a promising multiple access scheme.


For downlink RSMA transmission, taking 1-layer rate splitting as an example, all users' messages are split into a common part and a private part. The common parts are then combined and coded into a common stream, while the private parts are coded into separate private streams for each user. The common stream and private streams are superimposed at the transmitter side. Using a shared codebook, each user can decode the common stream first, treating the private streams as noise. Then, the user can subtract the common stream from the received signal using successive interference cancellation (SIC) and finally decode their private stream by treating the other private streams as noise \cite{ref1}. While downlink RSMA is a well-researched topic, the uplink has not been explored extensively. Uplink RSMA does not distinguish the common and private streams. Instead, the user's information is split into two parts and encoded independently, as if the user is divided into two virtual users. The base station (BS) then uses SIC technology to decode all user information in a specific decoding order \cite{ref2}.

In existing works on uplink RSMA, the authors in \cite{ref3} studied the performance of two-user uplink RSMA in terms of outage probability and achievable sum rate. In \cite{ref4}, fixed rate splitting (FRS) and cognitive rate splitting (CRS) for two-user uplink are studied to improve user fairness and outage performance, and a closed-form expression for the outage probability is given. The authors in \cite{ref5} investigated reconfigurable intelligent surface (RIS)-assisted two-user uplink RSMA, and optimized the transmit power and beamforming to maximize the achievable rate. The research \cite{ref6} also derived a closed-form expression for the outage probability of two-user uplink RSMA and used it to derive the expression for user throughput. Cooperative non-orthogonal multiple access (C-NOMA) and cooperative RSMA (C-RSMA) for two-user uplink were proposed in \cite{ref7}, with consideration of the proportional fairness coefficient. The aforementioned studies were conducted with two-user uplink RSMA. However, the complexity of obtaining the optimal decoding order for the uplink RSMA increases exponentially as the user number rises. The authors in \cite{ref8} proposed a suboptimal decoding order based on channel gain and message splitting ratio. Meanwhile, \cite{ref9} adopted an exhaustive search method to find the optimal decoding order. Despite these efforts, there remains a need for an effective approach to obtain the optimal decoding order.

While many studies focus on metrics such as user fairness, sum rate, or outage probability, there are some automatic control situations where latency is a critical parameter. One example is vehicle platooning, where the lead vehicle must wait for movement status updates from all platooning vehicles before making control decisions. This highlights the importance of minimizing latency for all users. In addition, the achievable latency and complexity of existing algorithms should be further reduced. In summary, our main contributions are as follows:

\begin{itemize}
  \item Unlike existing works, we propose an advanced uplink RSMA technique that considers each user's packet length and minimizes the maximum latency for all users. This enables fast and efficient control decisions in automated control scenarios.
  \item The complexity of obtaining the optimal decoding order in the formulated optimization problem increases exponentially with the number of users. To reduce the complexity, we propose an user pairing scheme and transform the optimization for power and decoding order in the original problem into optimization for power and bandwidth.
  \item Through theoretical analysis, the closed-form expressions for power and bandwidth allocation for a given latency are first derived. Then a bisection method is used to determine the minimum latency and optimal resource allocation.
  \item Numerical simulations are performed to verify the superior performance of our proposed algorithm. The results show that our proposed algorithm outperforms NOMA. Moreover, the proposed algorithm outperforms unpaired RSMA with suboptimal decoding order \cite{ref8} and approaches the performance of unpaired RSMA with exhaustive method \cite{ref9} when the number of users is small. In addition, our proposed algorithm significantly reduces the computational complexity.
\end{itemize}

\section{SYSTEM MODEL AND PROBLEM FORMULATION}
Without loss of generality, this letter considers a single input single output (SISO) uplink RSMA transmission, which contains a BS and $N$ users. The channel gain from the user $n$ to BS is denoted by $h_{n}$. Using RSMA, the information $x_n$ of user $n$ is divided into $x_{n1}$ and $x_{n2}$. The message received by BS can be expressed as $y_{BS}=\sum^{N}_{n=1}\sum^{2}_{j=1}{\sqrt{h_{n}}\cdot x_{nj}}+n_{0}$, where $n_{0}$ denotes additive Gaussian white noise. After SIC is performed at BS, the rate of $x_{nj}$ can be expressed as $r_{nj}=B\log_{2}(1+\frac{h_np_{nj}}{\sum_{\{(n'\in N,j'\in J)\mid\pi_{n'j'}>\pi_{nj}\}}h_{n'}p_{n'j'}+\sigma^{2}B})$, where $B$ is the bandwidth, $p_{nj}$ is the transmission power allocated to $x_{nj}$, $\pi_{nj}$ is the decoding order of $x_{nj}$, and $\sigma^{2}$ is the power spectral density of the Gaussian noise. So the transmission rate of user $n$ can be expressed as $r_{n}=\sum^2_{j=1}r_{nj}$, and the transmission latency can be write as $t_n={PL}_n/r_n$, where ${PL}_n$ is the package length of user $n$.


In some automatic control systems, the BS needs to obtain information from all users before making control decisions, so we build the following optimization problem to minimize the maximum transmission latency.
\begin{align}\label{eq1}
\boldsymbol{P1}: \hspace{10mm}  &\min_{\boldsymbol{\pi}\boldsymbol{p}}\max\quad t_n \\
s.t.  \ \ &\  p_{nj}>0, \quad 1\leq n \leq N, \quad j\in[1,2];\tag{1a}\\
            &\ \boldsymbol{\pi} \in\mathbf{\Pi};  \tag{1b}\\
            &\  \sum\nolimits^{2}_{j=1} p_{nj}\leq P^{max}_n, \quad 1\leq n \leq N. \tag{1c}
\end{align}
where permutation $\boldsymbol{\pi}$ is the decoding order of the BS, $\boldsymbol{p}=[p_{11},p_{12},p_{21},...,p_{n2}]$, $\mathbf{\Pi}$ denotes all possible decoding orders and $P^{max}_n$ is the maximum transmission power of user n.

Using $\tau$ to denote the upper bound of latency for all users, we have $t_n\leq \tau, 1\leq n \leq N$, Then problem $\boldsymbol{P1}$ can be transformed into:
\begin{align}\label{eq2}
\boldsymbol{P2}: \hspace{10mm}  &\min_{\boldsymbol{\pi}\boldsymbol{p}} \tau \\
s.t.  \ \ &\  r_{n}\geq{{PL}_n}/{\tau}, \quad 1\leq n \leq N;\tag{2a}\\
            &\  p_{nj}>0, \quad1\leq n \leq N, \quad j\in[1,2];\tag{2b}\\
            &\ \boldsymbol{\pi}\in\boldsymbol{\Pi};  \tag{2c}\\
            &\  \sum\nolimits^{2}_{j=1} p_{nj}\leq P^{max}_n, \quad 1\leq n \leq N; \tag{2d}
\end{align}

In problem $\boldsymbol{P2}$, the decoding order $\boldsymbol{\pi}$ is a discrete variable. Condition (2a) is in the form of the sum of two logarithmic functions. To solve this problem, we can exhaust $\boldsymbol{\pi}$ and use the successive convex approximation (SCA) algorithm to transform condition (2a) into a convex form. However, the decoding order set $\boldsymbol{\Pi}$ contains $(2N)!/2^{N}$ elements, which makes the exhaustive method highly complex as the number of users increases. Since existing studies have determined the optimal decoding order for two users\cite{ref9}, we consider pairing every two users to reduce the complexity.

The performance of pairing in NOMA is determined by channel gain\cite{ref10}. Building on this finding, authors in \cite{ref9} compared three pairing methods (strong-strong, strong-weak, and strong-middle) in the uplink RSMA scenario. The results indicated that the strong-weak pairing scheme performed the best. In our proposed algorithm, the packet length of each user is considered. However, simulation tests show that the impact of packet length on pairing performance is negligible compared to channel gain. Therefore, this letter uses the strong-weak pairing strategy, where the user with the strongest channel condition is paired with the weakest. The bandwidth resources are allocated orthogonally between different pairs. Upon receiving resource requests from users, the BS sorts the channel gains of all users and generates a pairing table by the strong-weak pairing strategy. Subsequently, the BS generates the resource allocation results for each user according to the proposed algorithm and returns the results to the users.

Suppose there are $M$ pairs, and each pair contains two users. The channel gain of the $k$-th user in the $m$-th pair to the BS is denoted by $h_{k}^{m}, k\in[1,2], 1\leq m \leq M$. Research \cite{ref11} shows that uplink RSMA transmission of two users can achieve all rate regions by splitting the information of only one user. Without loss of generality, suppose the message of the first user in each pair is split into two parts $x^m_{11}$ and $x^m_{12}$ , and the message of the second user is not split, the optimal decoding order at the BS is $x^m_{11}$, $x^m_{2}$, $x^m_{12}$.

The rate of $x^m_{11}$, $x^m_{2}$ and $x^m_{12}$ can be expressed as follows:
\begin{equation}\label{eq3}
r^m_{11}=B\alpha_m \log_2(1+\frac{h^m_1 p^m_{11}}{h^m_2p^m_2+h^m_1p^m_{12}+\sigma^2B\alpha_m}),\\
\end{equation}
\begin{equation}\label{eq4}
r^m_{2}=B\alpha_m \log_2(1+\frac{h^m_2p^m_2}{h^m_1p^m_{12}+\sigma^2B\alpha_m}),\\
\end{equation}
\begin{equation}\label{eq5}
r^m_{12}=B\alpha_m \log_2(1+\frac{h^m_1p^m_{12}}{\sigma^2B\alpha_m}),\\
\end{equation}
where $\alpha_m$ is the bandwidth allocation factor with $\sum_{m=1}^{M}\alpha_m\leq1$, $p^m_{11}$, $p^m_{2}$ and $p^m_{12}$ are the power allocated to $x^m_{11}$, $x^m_{2}$ and $x^m_{12}$ respectively. So the rate of user 1 in $m$-th pair is $r^m_1=r^m_{11}+r^m_{12}$ and the transmission latency of user $k$ in the $m$-th pair is $t^m_k=PL^m_k/r^m_K$, where ${PL}^m_k$ is the package length of user $k$ in the $m$-th pair.

The optimization of the power and decoding order in $\boldsymbol{P2}$  can be translated into the optimization of the bandwidth and power allocation in $\boldsymbol{P3}$  as follows:
\begin{align}\label{eq6}
\boldsymbol{P3}: \hspace{10mm}  &\min_{\boldsymbol{\alpha}\boldsymbol{p}}\quad \tau \\
s.t.  \ \   &\ r^m_{k}\geq{PL^m_k}/{\tau}, 1\leq m \leq M, k\in[1,2];\tag{6a}\\
            &\  p^m_{kj}>0, 1\leq m \leq M, k,j\in [1,2];\tag{6b}\\
            &\ \sum\nolimits^M_{m=1} \alpha_m \leq 1;  \tag{6c}\\
            &\  \sum\nolimits^{2}_{j=1} p^m_{kj}\leq P^m_{kmax}, 1\leq m \leq M, k,j\in[1,2]; \tag{6d}
\end{align}
where $\boldsymbol{\alpha}=[\alpha_1,\alpha_2,...,\alpha_M]$ is a vector of bandwidth allocation factors.

\section{RESOURCE ALLOCATION ALGORITHM FOR PAIRED RSMA}
Since the expression of transmission rate is a monotonically increasing function of power and bandwidth, we can use contradiction to prove that the optimal solution of $\boldsymbol{P3}$ is obtained when and only when all users have the same latency. in other words, the ratio of optimized rates is $r^1_1:r^1_2:r^2_1:...:r^M_2={PL}^1_1:{PL}^1_2:{PL}^2_1:...:{PL}^M_2$. Thus, the rates of two users in the same pair have $r^m_2=\frac{{PL}^m_2}{{PL}^m_1}r^m_1$.

\begin{figure}[!t]
\centering
\includegraphics[width=2.0in]{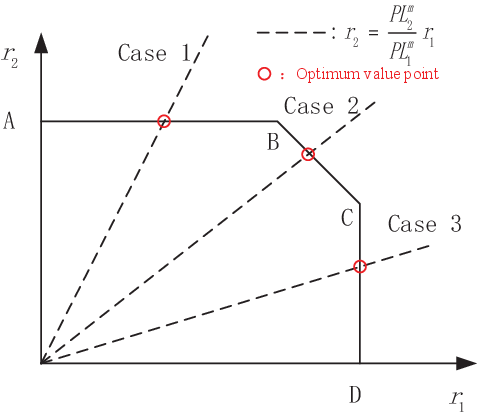}
\caption{Rate region of $r_1$ and $r_2$ in the same pair in the uplink RSMA}
\label{fig:1}
\end{figure}

Fig. \ref{fig:1} shows the achievable rate region for two users in the uplink RSMA. When the decoding order is $x^m_{11}$, $x^m_{2}$ and $x^m_{12}$, all points on the rate region can be reached.

For the line AB, the power allocation is:
\begin{equation}\label{eq7}
0\leq p^m_{11}\leq P^m_{1max},\quad p^m_{12}=0,\quad p^m_{2}=P^m_{2max},\\
\end{equation}
User 2 has the maximum rate as:
\begin{equation}\label{eq8}
 r^m_{2}=B\alpha_m \log_2(1+\frac{h^m_2P^m_{2max}}{\sigma^2B\alpha_m}).
\end{equation}

For the line BC, the power allocation is:
\begin{equation}\label{eq9}
p^m_{11}+ p^m_{12}=P^m_{1max},\quad p^m_{2}=P^m_{2max},\\
\end{equation}
The sum rate of the two user is:
\begin{equation}\label{eq10}
r^m_{1}+r^m_{2}=B\alpha_m \log_2(1+\frac{h^m_1P^m_{1max}+h^m_2P^m_{2max}}{\sigma^2B\alpha_m}).
\end{equation}

For the line CD, the power allocation is:
\begin{equation}\label{eq11}
p^m_{11}=0,\quad p^m_{12}=P^m_{1max},\quad 0\leq p^m_{2}\leq P^m_{2max},\\
\end{equation}
User 1 has the maximum rate as :
\begin{equation}\label{eq12}
r^m_{1}=B\alpha_m \log_2(1+\frac{h^m_1P^m_{1max}}{\sigma^2B\alpha_m}).
\end{equation}

According to the above analysis, the optimal power allocation of problem $\boldsymbol{P3}$ is the intersection of line $r^m_2=\frac{{PL}^m_2}{{PL}^m_1}r^m_1$ and the rate region. As shown in Fig. \ref{fig:2}, the intersection point may exist three cases.

Case 1: The intersection point is on AB, and according to (\ref{eq3}), (\ref{eq7}), (\ref{eq8}) and (6a), we have:
\begin{equation}\label{eq13}
B\alpha^{AB}_m \log_2(1+\frac{h^m_2P^m_{2max}}{\sigma^2B\alpha^{AB}_m})=\frac{{PL}^m_2}{\tau},
\end{equation}
\begin{equation}\label{eq14}
p^m_{11}=\frac{(2^\frac{{PL}^m_1}{\tau B\alpha^{AB}_m}-1)(h^m_2P^m_2max+\sigma^2B\alpha^{AB}_m)}{h^m_1}.
\end{equation}

Case 2: The intersection point is on BC, and according to (\ref{eq4}), (\ref{eq9}), (\ref{eq10}) and (6a), we have:
\begin{equation}\label{eq15}
B\alpha^{BC}_m \log_2(1+\frac{h^m_1P^m_{1max}+h^m_2P^m_{2max}}{\sigma^2B\alpha^{BC}_m})=\frac{{PL}^m_1+{PL}^m_2}{\tau},
\end{equation}

\begin{equation}\label{eq16}
p^m_{12}=\frac{h^m_2P^m_{2max}}{h^m_1(2^\frac{{PL}^m_2}{\tau B\alpha^{BC}_m}-1)}-\frac{\sigma^2B\alpha^{BC}_m}{h^m_1},
\end{equation}
The $r^m_{12}$ can be obtained by taking $p^m_{12}$ into (\ref{eq5}), then $r^m_{11}={PL}^m_1/ \tau- r^m_{12}$, and according to (\ref{eq3}) we can obtain $p^m_{11}$:
\begin{equation}\label{eq17}
p^m_{11}=\frac{(2^\frac{r^m_{11}}{B\alpha^{BC}_m}-1)(h^m_2P^m_{2max}+h^m_1p^m_{12}+\sigma^2B\alpha^{BC}_m)}{h^m_1}.
\end{equation}

Case 3: The intersection point is on CD, and according to (\ref{eq11}), (\ref{eq12}) and (6a):
\begin{equation}\label{eq18}
B\alpha^{CD}_m \log_2(1+\frac{h^m_1P^m_{1max}}{\sigma^2B\alpha^{CD}_m})=\frac{{PL}^m_1}{\tau},
\end{equation}
\begin{equation}\label{eq19}
p^m_{2}=\frac{(2^\frac{{PL}^m_2}{\tau B\alpha^{CD}_m}-1)(h^m_1P^m_1max+\sigma^2B\alpha^{CD}_m)}{h^m_2}.
\end{equation}

According to (\ref{eq13}), (\ref{eq15}) and (\ref{eq18}), the expression of $\alpha^{AB}_m$,$\alpha^{BC}_m$ and $\alpha^{CD}_m$ are given in (\ref{eq20}), (\ref{eq21}) and (\ref{eq22}).
\begin{figure*}[ht] 
	\begin{equation}\label{eq20}
	\alpha^{AB}_m = \frac{-\ln2h^m_2P^m_{2max}{PL}^m_2}{\tau Bh^m_2P^m_{2max}W(\frac{-\ln2{PL}^m_2\sigma^2}{\tau h^m_2P^m_{2max}}e^{\frac{-\ln2{PL}^m_2\sigma^2}{\tau h^m_2P^m_{2max}}})+\ln2{PL}^m_2\sigma^2B},
	\end{equation}
	\begin{equation}\label{eq21}
\alpha^{BC}_m = \frac{-\ln2(h^m_1P^m_{1max}+h^m_2P^m_{2max})({PL}^m_1+{PL}^m_2)}{\tau B(h^m_1P^m_{1max}+h^m_2P^m_{2max})W(\frac{-\ln2({PL}^m_1+{PL}^m_2)\sigma^2}{\tau (h^m_1P^m_{1max}+h^m_2P^m_{2max})}e^{\frac{-\ln2({PL}^m_1+{PL}^m_2)\sigma^2}{\tau (h^m_1P^m_{1max}+h^m_2P^m_{2max})}})+\ln2({PL}^m_1+{PL}^m_2)\sigma^2B},
	\end{equation}
	\begin{equation}\label{eq22}
	\alpha^{CD}_m = \frac{-\ln2h^m_1P^m_{1max}{PL}^m_1}{\tau Bh^m_1P^m_{1max}W(\frac{-\ln2{PL}^m_1\sigma^2}{\tau h^m_1P^m_{1max}}e^{\frac{-\ln2{PL}^m_1\sigma^2}{\tau h^m_1P^m_{1max}}})+\ln2{PL}^m_1\sigma^2B},
	\end{equation}
\end{figure*}
Where $W(\cdot)$ is the Lambert-W function which satisfies $W(xe^x)=x$. Note that the Lambert-W function may have multiple solutions, and the appropriate solution should be chosen.

With the closed-form expressions for bandwidth allocation and power allocation, we can solve the problem $\boldsymbol{P3}$ by the bisection method. For each given $\tau$, a set of $\alpha^{AB}_m$, $\alpha^{BC}_m$ and $\alpha^{CD}_m$ can be calculated according to (\ref{eq20}), (\ref{eq21}) and (\ref{eq22}), and the power allocation corresponding to each case can be obtained by (\ref{eq14}), (\ref{eq16}), (\ref{eq17}) and (\ref{eq19}), and then the condition (6d) is used to judge which case occurs. The specific algorithm is shown in Algorithm 1.
\begin{algorithm}\small
   \caption{User pairing-based power allocation algorithm.}
   \begin{algorithmic}[1]
   \STATE {Initialize upper and lower bound $\tau_{ub}$, $\tau_{lb}$, tolerance $\varepsilon$.}
   \WHILE{ $\tau_{ub}-\tau_{lb}>\varepsilon$ }
       \STATE set   $ \tau={(\tau_{ub}+\tau_{lb})}/{2}$
       \FOR{m=1:$M$}
       \STATE calculate $\alpha^{AB}_m$, $\alpha^{BC}_m$ and $\alpha^{CD}_m$ respectively according to (\ref{eq20}), (\ref{eq21}) and (\ref{eq22}).
       \STATE Calculate the power allocation for each case according to (\ref{eq14}), (\ref{eq16}), (\ref{eq17}) and (\ref{eq19}) respectively.

       \IF {The power allocation of case 1 satisfies (6d)}
           \STATE $\alpha_m=\alpha^{AB}_m$.

       \ELSIF {The power allocation of case 2 satisfies (6d)}
            \STATE $\alpha_m=\alpha^{BC}_m$.
       \ELSIF {The power allocation of case 3 satisfies (6d)}
            \STATE $\alpha_m=\alpha^{CD}_m$.
            \ELSE
            \STATE set $\tau_{lb}=\tau$, break and jump to step 2.
       \ENDIF
       \ENDFOR
        \IF {$\sum_{m=1}^{M}\alpha_m\leq1$}
        \STATE set $\tau_{ub}=\tau$.
        \ELSE
        \STATE set $\tau_{lb}=\tau$.
        \ENDIF
   \ENDWHILE
   \STATE {Output $\tau$, $\alpha_m$, $p^m_{11}$, $p^m_{12}$, and $p^m_{2}$.}
   \end{algorithmic}
   \end{algorithm}

The complexity of Algorithm 1 in each iteration is to check which case satisfies the power constraint (6d), which introduces the complexity of $ \mathcal{O}(M)$ according to (\ref{eq14}), (\ref{eq16}), (\ref{eq17}) and (\ref{eq19})-(\ref{eq22}). In addition the complexity of the bisection method with accuracy $\varepsilon$ is $ \mathcal{O}(\log_2(1/\varepsilon))$\cite{ref12}, so the total complexity of Algorithm 1 is $ \mathcal{O}(M\log_2(1/\varepsilon))$.

\section{SIMULATION RESULT AND DISCUSSIONS}
In this section, we set $N$ users uniformly distributed within a radius of 200 m from the BS. The path loss model is $128.1+37.6\log_{10}d$ ($d$ is in km). The total bandwidth set to 1 MHz, and the noise power spectral density is $\sigma^2=-174 $ dBm/Hz \cite{ref9}. Each user has the same power limit $P_{max}$ and randomly generates a packet of 50-1200 bytes. The algorithms for the comparison include unpaired RSMA using an exhaustive method to obtain the optimal decoding order \cite{ref9}, unpaired RSMA using a suboptimal decoding order \cite{ref8}, paired NOMA \cite{ref10} and unpaired NOMA. Without loss of generality, the two unpaired RSMA algorithms assume that all users' messages are split into two parts. All simulation were performed on an Intel Core i9-13900KF CPU @ 5.8 GHz and 32G RAM using MATLAB R2023a.
\begin{figure}[!t]
\centering
\includegraphics[width=2.5in]{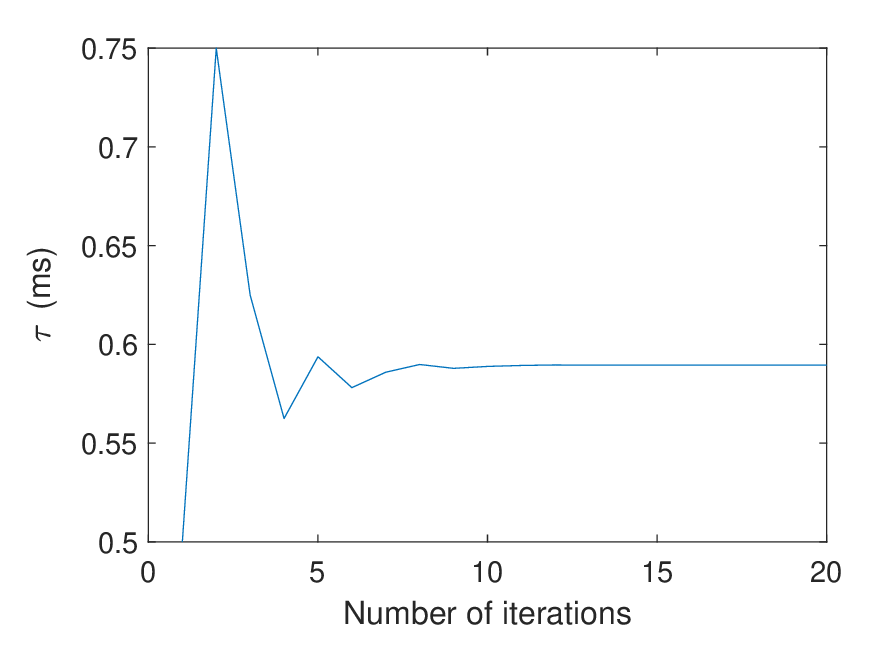}
\caption{Convergence process of the proposed paired RSMA when the number of users $N=4$.}
\label{fig:2}
\end{figure}

Fig. \ref{fig:2} illustrates the convergence process of the proposed algorithm. The maximum latency of all users in each iteration is denoted by $\tau$, and convergence is observed after ten iterations.
\begin{figure}[!t]
\centering
\includegraphics[width=2.5in]{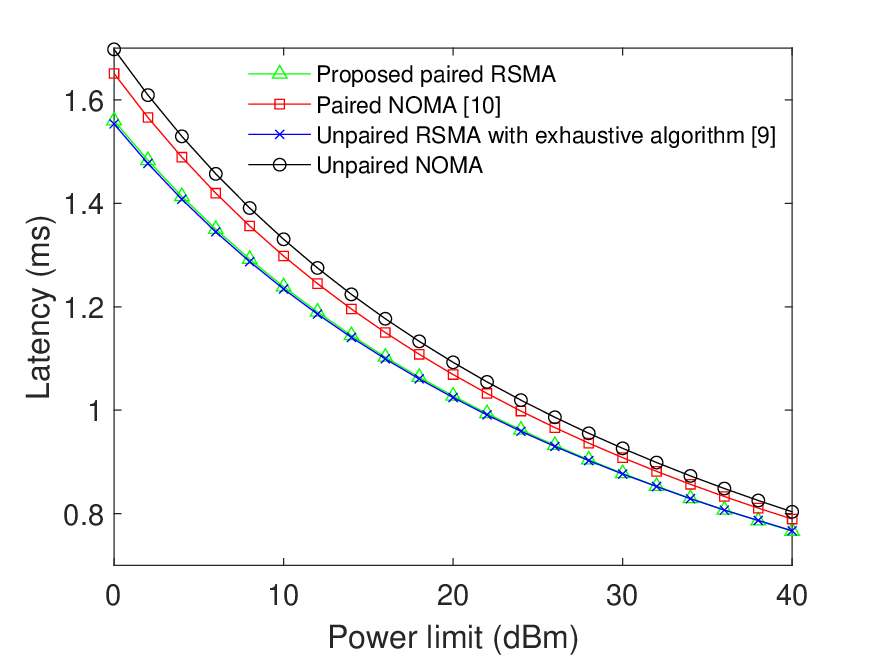}
\caption{Latency performance with different power limits when the number of users $N=4$.}
\label{fig:3}
\end{figure}

Fig. \ref{fig:3} simulates the transmission latency of all schemes for four users ($N=4$) at different power limits. In this case, RSMA uses an exhaustive method to obtain the optimal decoding order. As the maximum power $P_{max}$ increases, the latency decreases for all schemes. It can be seen that RSMA always outperforms NOMA. The performance of the proposed paired RSMA is similar to the unpaired RSMA, while paired NOMA outperforms unpaired NOMA. In addition, both RSMA and NOMA show a convergence of unpaired and paired at high power. The reason is that unpaired RSMA can achieve every point of the rate region \cite{ref11}. The boundary of the rate region is the theoretical upper bound of RSMA, and the proposed paired RSMA can approach this theoretical upper bound with a low-complexity algorithm. NOMA cannot achieve all the points on the rate region. User pairing makes the resource allocation of NOMA more flexible and reduces the interference of the SIC process, which leads to some gains.


\begin{figure}[!t]
\centering
\includegraphics[width=2.5in]{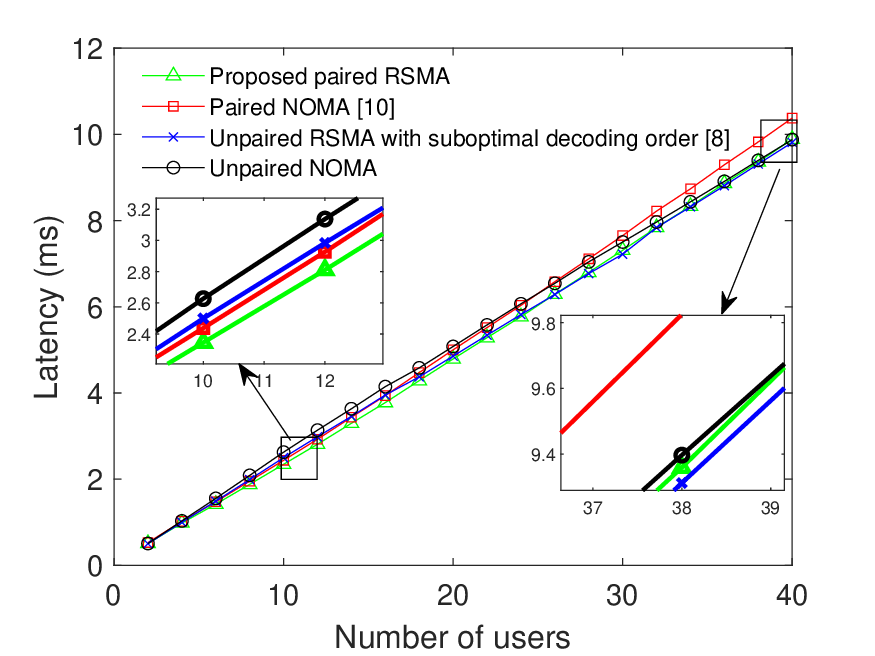}
\caption{Latency performance for different total number of users when the power limit $P_{max}=23$ dBm.}
\label{fig:4}
\end{figure}

The computational complexity of the exhaustive algorithm grows exponentially as the total number of users increases. In order to compare latency performance at different total numbers of users, we adopt a suboptimal decoding order \cite{ref8} rather than the exhaustive method. Fig. \ref{fig:4} shows that RSMA outperforms NOMA, whether paired or unpaired. Different from Fig. \ref{fig:3}, the proposed paired RSMA outperforms the unpaired RSMA with suboptimal decoding order for the number of users $N \leq 24$. However, when $N > 24$, the performance of RSMA with suboptimal decoding order will exceed the performance of the proposed paired RSMA due to the fact that more number of users means more number of pairs and therefore less bandwidth is allocated to each pair.

\begin{table}
\begin{center}
\caption{Comparison of the computational complexity of each scheme}
\label{tab1}
\setlength{\tabcolsep}{1mm}{
\begin{tabular}{| c | c | c | c | c | c |}
\hline
\multirow{2}*{Schemes} &\multicolumn{5}{c|}{Simulation time}\\
\cline{2-6}
~ & $N=4$ & $N=10$ & $N=20$& $N=30$ & $N=40$ \\
\hline
\makecell[c]{Proposed \\paired RSMA}& 0.068 s&0.095 s&0.106 s&0.141 s&0.161 s \\
\hline
\makecell[c]{Unpaired \\RSMA with \\exhaustive method}&3447.5 s & -& - & - & -\\
\hline
\makecell[c]{ Unpaired RSMA\\ with suboptimal \\decoding order}&1.783 s&7.378 s&19.002 s&36.117 s&74.457 s\\
\hline
\makecell[c]{Paired NOMA}&4.084 s&4.515 s&5.737 s&7.589 s&10.113 s\\
\hline
\makecell[c]{Unpaired NOMA}&4.031 s&4.260 s&4.603 s&5.106 s&5.271 s\\
\hline
\end{tabular}}
\end{center}
\end{table}

To further evaluate the benefits of the proposed algorithm in terms of complexity, Table \ref{tab1} gives the simulation time for each scheme at different total numbers of users. For unpaired RSMA, the computational complexity introduced by the exhaustive decoding order method is unacceptable, and even given a suboptimal decoding order, the computational complexity grows rapidly when the number of users increases. On the contrary, the proposed paired RSMA algorithm gives the closed-form expression for resource allocation, achieving similar performance to unpaired RSMA with significantly lower complexity.

Based on the results, it is clear that RSMA outperforms NOMA. In addition, the proposed algorithm significantly reduces the computational complexity. It outperforms unpaired RSMA with suboptimal decoding order \cite{ref8} and approaches the performance of unpaired RSMA with exhaustive method \cite{ref9} when the number of users is small. As the number of users increases, the performance of the proposed paired RSMA is limited by the bandwidth, which requires a trade-off between performance and complexity.

\section{Conclusion}
In this letter, to address the latency problem in some automatic control scenarios, we propose a resource allocation algorithm for uplink RSMA to minimize the maximum transmission latency. However, the complexity of uplink RSMA to confirm the optimal decoding order grows exponentially with the total number of users, so user pairing is introduced and a low-complexity resource allocation algorithm is designed to transform the optimization of power and decoding order into the optimization of power and bandwidth allocation. The achievable rate region of two-user uplink RSMA is first analyzed, and the closed-form expressions for power and bandwidth allocation for a given latency are derived. Then a bisection method is used to determine the minimum latency and optimal resource allocation. Results show that the proposed low-complexity algorithm based on user pairing significantly reduces the computational complexity and achieves similar performance to unpaired uplink RSMA.




\newpage


\vfill

\end{document}